# Non-Closed Acoustic Cloak Enabled by Sequential-Step Linear Coordinate Transformations

Z. Basiri, MH. Fakheri and A. Abdolali [*]

*Applied Electromagnetic Laboratory, School of Electrical Engineering, Iran University of Science and Technology,*

*Tehran, 1684613114, Iran*

* abdolali@iust.ac.ir

**Abstract**

Hitherto acoustic cloaking devices, which conceal objects externally, have depended on the objects' characteristics. Despite previous works, we design cloaking device placed neighbor an arbitrary object and makes it invisible without the need to make it enclosed. Applying sequential linear coordinate transformations, leads to a non-closed acoustic cloak (NCAC) with homogeneous materials that creates an open invisible region. We propose a non-closed carpet cloak to conceal objects on a reflecting plane. Numerical simulations verify the cloaking effect, which is completely independent of the geometry and material properties of the hidden object. Due to the simple acoustic constitutive parameters of presented structures, this work paves the way toward realization of non-closed acoustic devices, which could find applications in air born sound manipulation and underwater demands.

**Introduction**

Invisibility cloak is one of the most attractive research topics due to its exotic properties in deflecting the waves around the objects. A powerful approach to achieve invisibility devices, is the coordinate transformation method that firstly utilized by Pendry et.al for design of electromagnetic (EM) ideal cloak [1]. Not long after, Cummer and Schurig presented a two dimensional (2D) acoustic cloak by illustrating the analogy between 2D time harmonic Maxwell equations and acoustic wave equations [2]. Later, a similar coordinate transformation method further extended to design of 3D acoustic cloaks by Chen and Chan [3]. The same relations through acoustic scattering theory also derived in [4] by Cummer et.al.. Due to the potential applications of cloaking in variety of both acoustics and electromagnetics scenarios, numerus schemes have been devoted to theoretically development and fabrication of cloaking devices [5-10]. The major difficulty of ideal cloak implementation, is the requirement of extreme material properties which is resulted from the transformation of a 0D point to a 2D cloaked region. To get rid of this obstacle, the concept of carpet cloak was proposed [11]. Carpet cloak or ground plane cloak, is a device that restore the signature of the target as if the incident wave reflected from a mirror plane. The carpet cloak, is designed by employing a coordinate transformation from a 2D flat line segment to a 2D curved line that creates a bump with the mirror plane, which resolves the need to extreme materials' parameters. Later, inspired by carpet cloak concept, unidirectional free space cloak which is designed to represent the cloaking effect for a specified direction of propagation, was proposed [12]. The strategy to unidirectional cloak design is based on the property of the mirror plane to be invisible when probed by a plane wave propagating parallel to it.

The first proposal of carpet cloak and unidirectional cloak was presented through quasi-conformal mapping [11, 12] with remarkable advantage of minimizing the anisotropy of obtained materials [12-14]. However, inhomogeneous structure of quasi-conformal cloaks, leads to a difficult fabrication process and neglecting the weak anisotropy, gets a lateral shift in the reflected wave [15]. Another disadvantage is the size of this type of cloaks, which is bulky compared to that of the target. To overcome this challenges, linear transformation based carpet cloak [16-19] and unidirectional cloak [20-23] were proposed. Cloaks based on linear coordinate transformation have homogeneous constitutive parameters with finite anisotropy which obviates the need to space dependent materials. Till now, besides huge amount of theoretical investigations to advance of cloaks [24-28], several researches experimentally demonstrated acoustic carpet cloaks via homogeneous fluid like materials. For example, perforated plastic plates have been applied for 2D [29] and 3D [30] ground plane cloaks in air-borne acoustics and steal strips [31] or brass plates [32] have been proposed for underwater acoustics. Acoustic unidirectional cloaks also were implemented via composites of metals and porous materials for multi-layered host medium [33] and meta-fluid structured by slab-shape units for air host [34].

Although considerable progress has been made in invisibility cloaks, all conventional cloaking devices are "interior" cloaks and prevent the target from interact with outer world that is a restriction for the usage of these applicable devices. In order to obviate this drawback, Lai et.al [35] presented "external "cloaking and illusion devices for EM framework. Because of relevant acoustic demands, not long after the first proposal, the idea was extended to acoustics by Zhu et.al [36]. The key enabling feature of designing such external or "non-closed" invisibility devices, is complementary media concept, which is designed to conceal a predefined object [35]. However, the complementary media based devices depend on the shape and material properties of the hidden object. Therefore, any changes in objects description or any movements disturb the cloaking effect. Despite the recent developments in object independent EM non-closed cloaking devices with simple materials [37, 38], acoustic non-closed cloaks are all designed based on complementary object method [39-42]. Therefore, the dependence of the cloaking device on target and inevitable inhomogeneity of the obtained materials are remaining challenges that make the implementation of conventional acoustic non-closed cloaking devices nonrealistic from the practical point of view.

Differ from conventional design, in this paper; we design non-closed acoustic cloak (NCAC), which its materials are feasibly independent of the target geometry and its constitutive materials. The proposed technique is based on applying sequential linear transformations to create a non-closed invisible region positioned on a reflecting plane. So any standing or moving object in the hidden region, will be acoustically invisible without being blinded. Due to the nature of linear coordinate transformation, the obtained materials of NCAC are homogeneous which results in easy to fabricate non-closed carpet cloak. Full wave numerical simulations using Finite Element Method verify the expected effect of NCAC.

**Results**

To start with a conception of idea, it is well understood that the conventional carpet cloak [17], has a hidden region underneath a sound hard boundary bump where any arbitrary object located there, becomes invisible. The main purpose is to design an acoustic cloak, which in addition to not enclosed the objects; its constitutive material is independent of the material, geometry and the position of them. Fig. 1 illustrates the equivalence between the behavior of desired non-closed device and its corresponded conventional one, where the reflected wave of the whole NCAC and the neighbor target is not disturbed.

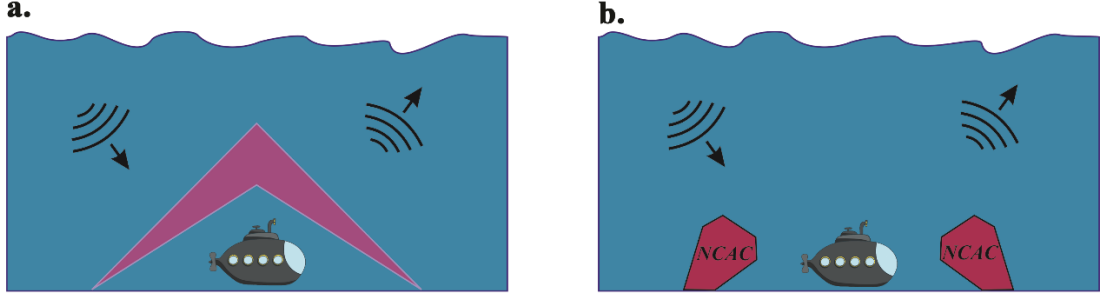

**Figure 1.** A schematic of (a) conventional carpet cloak and (b) NCAC for arbitrary object. It looks like the conventional one, with a created window in top of it. The desired non-closed carpet cloak has an acoustically equal signature with its corresponding conventional one. So, the incident wave is reflected from the non-closed device as a reflecting plane, without any perturbation.

At the first step, the material of conventional carpet cloak [17] will be derived. Without loss of generality, the problem is discussed in 2D framework while it could be extended to 3D case. As demonstrated in Figs. 2(a) and 2(b), the triangle $\triangle BCD$ in reference space $(x_0, y_0)$ is mapped to the quadrilateral BCDA in real space $(x_1, y_1)$. Therefore, any object located beneath the $BAD$ sound hard boundary (SHB) bump, becomes invisible.

As the second step, the existing SHB in the carpet cloak is physically eliminated by utilizing a linear folded transformation. The resulted folded medium is a type of complementary media [35] and is actually a SHB mimicking structure, without any imposed sound hard boundary condition.

For detail, considering the geometrical symmetry of the scheme, the problem is discussed only in the left half space of the reference and real spaces and a same method is applied to the right side. Analytically, a linear coordinate transformation that map $\triangle KBE$ region in reference space $(x_1, y_1)$ to $\triangle KBA$ region in real space $(x_2, y_2)$, folds the ground plane boundary $\overline{BE}$ to $\overline{AB}$ and makes an effective SHB on $\overline{AB}$ as shown in Fig. 2(c). Hence, the folded medium creates an illusionary SHB bump and regenerates the cloaking effect of conventional carpet cloak without making any disturbance in the scattered wave. From now on, for brevity, we call the domain mimics the sound hard boundary condition as "spoof sound hard boundary" (SSHB) and the carpet cloak without physical SHB bump as the carpet cloak with SSHBs.

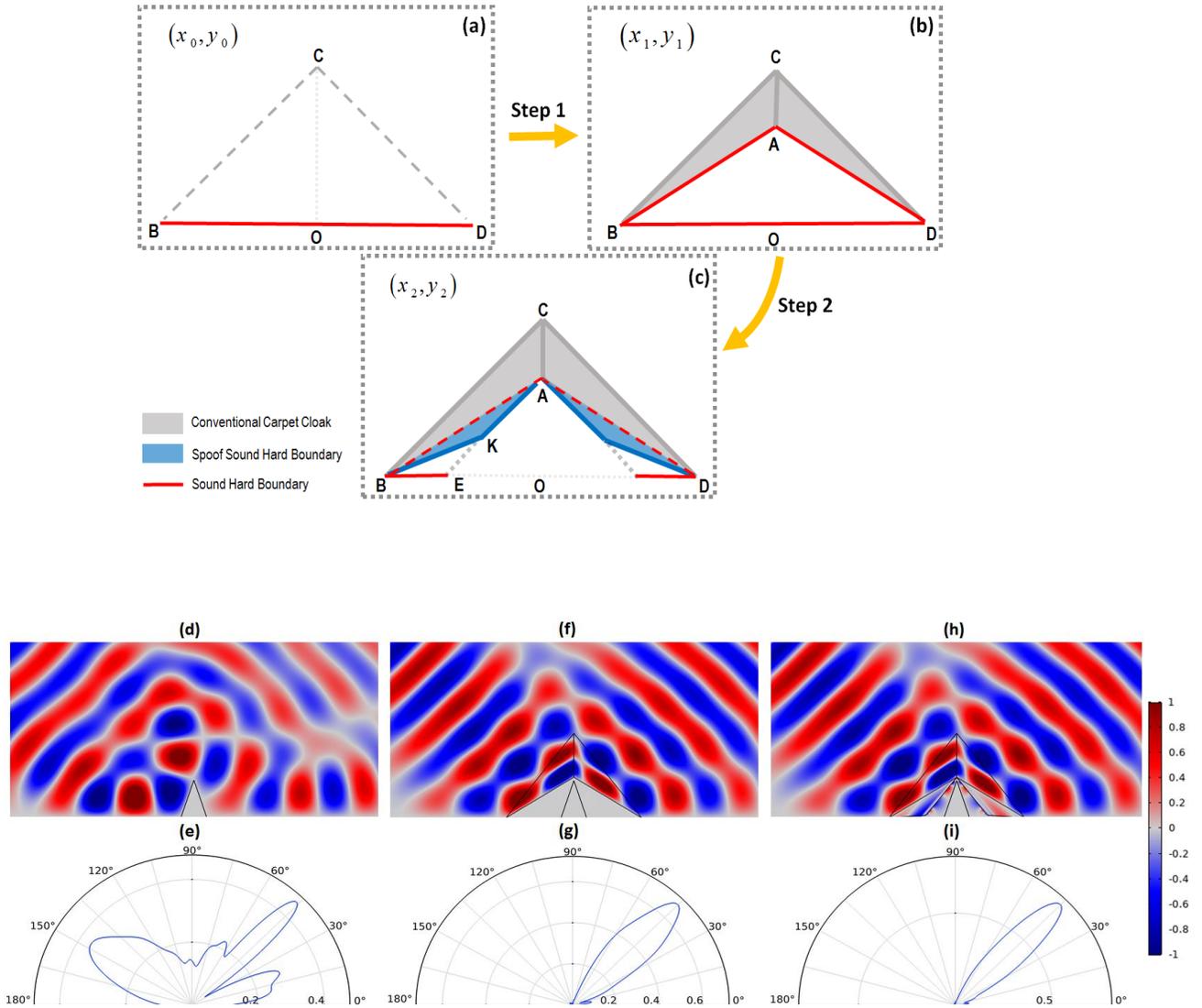

**Figure 2.** (a-c) Schematic diagram of first and second steps to NCAC design. (d-i) Comparison between the field distribution of target, conventional carpet cloak and first obtained transformation media, so called 'the carpet cloak with SSHBs'. (a) The reference space. (b) Conventional carpet cloak's scheme. (c) The spoof sound hard boundary folds the $\triangle KBE$ region in $(x_1, y_1)$ space to $\triangle KBA$ in $(x_2, y_2)$ space and makes an illusionary sound hard boundary condition on $\overline{AB}$ boundary that is shown by red dashed lines. (d-e) The near field distribution and far field scattering pattern of the object with sound hard boundaries. (f-g) The near field distribution and far field scattering pattern of conventional carpet cloak. (h-i) The near field distribution and far field scattering pattern of the carpet cloak with SSHBs with its obvious invisibility effect. It can be seen that the results in (f-g) and (h-i) are clearly similar to each other.

Finally, we can open a window in the carpet cloak with SSHBs to make its hidden region non-closed. Subsequently, as the third step, the carpet cloak with SSHBs and its surrounding fluid in reference space $(x_2, y_2)$ are compressed to smaller domains in the real space $(x_3, y_3)$ as shown in Fig. 3. For detail, a linear transformation that maps the $\Delta OCB$ region to the $\Delta OC'B$ as illustrated by red arrows in Fig. 3(c), compresses the carpet cloak with SSHBs to non-closed regions depicted by region 1. Similarly, the $\Delta BMC$ and $\Delta ENA$ regions of surrounding host fluid are also compressed to $\Delta BMC'$ and $\Delta ENA'$ domains denoted by region 2 in order to satisfy the matching condition [35, 43]. Thereupon, the compressing transformation, maps $\overline{NA}$, $\overline{AC}$ and $\overline{CM}$ boundaries in reference space $(x_2, y_2)$ to $\overline{NA'}$, $\overline{A'C'}$ and $\overline{C'M}$ boundaries in real space $(x_3, y_3)$.

In the coordinate transformation frame, by mapping the carpet cloak with SSHBs to the compressed regions, the path of the wave and outer boundaries of structure in the reference space, follow the compressing transformation in the real space and are mapped to transformed lines in compressed domains. In order to restore the path of wave, other folded regions are also utilized. The added domains, which are denoted by regions 3 and 4 shown in Fig. 3(c), are a type of complementary media [35] that convey the route of wave in real space to the compressed space by employing a linear transformation. The linear coordinate mapping, folds the black dashed lines representing the reference space to black solid lines in real space. The significant result of this mapping is the invariance of waves' path from reference space to the real space (Fig. 3(b)). Another remarkable implication of complementary regions, is mapping of each outer boundary to itself, which results in satisfaction of matching condition [35, 43]. Finally, all incident acoustic field is conveyed to the compressed domains and tracks the compression direction, bypasses the cloaked region and is scattered as the same with that in reference space.

To mathematically restate the complementary regions, firstly, the complementary medium $GHA'C'$ denoted by region 3 in Fig. 3(c), is obtained by mapping $\Delta OGC$ in the reference space to $\Delta OGC'$ in the real space. In fact, region 3 folds the $\overline{AC}$ boundary in reference space to $\overline{A'C'}$ in real space and $\overline{GH}$ boundary to itself as illustrated by yellow arrows in Fig. 3(c). Similarly, the complementary media of the compressed surrounding host fluid, denoted by region 4 in Fig. 3(c), are achieved by mapping $\Delta GMC$ and $\Delta HNA$ in the reference space, respectively to $\Delta GMC'$ and $\Delta HNA'$ in the real space. The obtained material of $\Delta GMC'$ region, folds the $\overline{GC}$ to $\overline{GC'}$ and $\overline{GM}$ to itself. In the same manner, also the material of $\Delta HNA'$ region, folds $\overline{NA}$ to $\overline{NA'}$ and $\overline{HN}$ to itself. In summary, the employed linear coordinate transformations to achieve the complementary materials drawn by region 3 and 4, are chosen in a way that $\overline{MC}$, $\overline{NA}$ and $\overline{AC}$ dashed line boundaries in reference space, respectively fold to $\overline{MC'}$, $\overline{NA'}$ and $\overline{A'C'}$ solid line boundaries in real space and all outer boundaries of real structure, i.e.

$\overline{MG}$, $\overline{GH}$ and $\overline{HN}$ are mapped to themselves (Fig. 3(c)). As the result, the compressed structure in the presence of complementary materials, is matched to host medium. Therefore, the compounded compressed regions and complementary materials have an identical scattering effect with the conventional carpet cloak in the reference space or equally, with a reflecting plane.

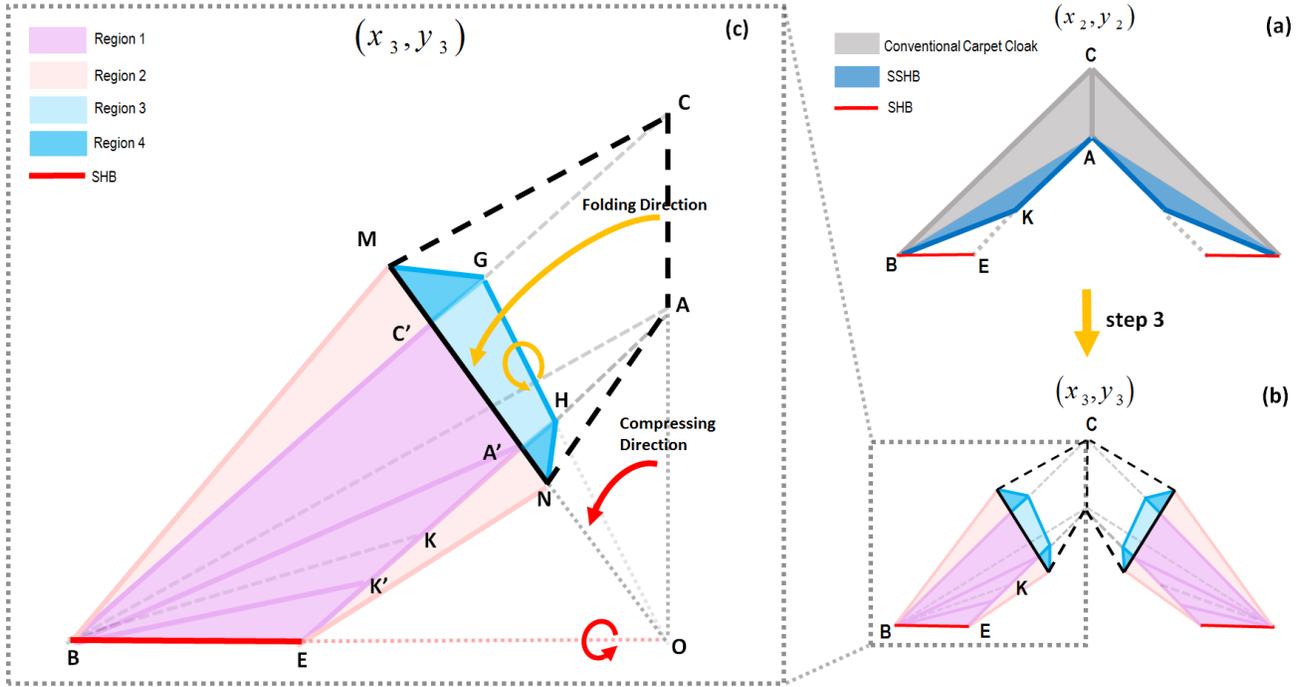

**Figure 3.** Schematic diagram of proposed object independent non-closed carpet cloak. (a) Conventional carpet cloak compressing. (b) All the $AEBC$ trapezoid region, is compressed to the smaller purple trapezoid region denoted by region 1 and the surrounding host medium, is also compressed to region 2. Blue domains, show the complementary materials. (c) A magnified view of the non-closed cloaks' architecture. The black dashed line boundaries, represent the original boundaries before compressing. Incident wave that meeting the dashed lines, is routed to the solid lines by complementary domains denoted by 3 and 4 regions. Significant result of this transformation, is that the signature of created window on conventional carpet cloak is canceled. Therefore, the whole structure is matched to host fluid and acts like an illusionary single reflecting plane.

In order to provide more intuitive perception of the designed NCAC, numerical simulations are performed using COMSOL Multiphysics finite element simulator. All simulations are carried out by adopting the acoustic pressure field in the frequency of 3.34 kHz and the host medium is chosen as air with $\rho_0 = 1 \ kg/m^3$ and $c_0 = 334 \ m/s$ values for mass density and speed of sound, respectively. Taking the left side of problems' geometry as the reference, at the first step of design, the acoustic constitutive materials for conventional cloaks' material and the SSHB region, are

$$\bar{\bar{\rho}}_{cloak} = \rho_0 \begin{bmatrix} 1.2871 & -1.2146 \\ -1.2146 & 1.9231 \end{bmatrix}, \kappa_{cloak} = 1.9231\kappa_0 \text{ and } \bar{\bar{\rho}}_{SSHB} = \rho_0 \begin{bmatrix} -16.0124 & 13.0663 \\ 13.0663 & -10.7247 \end{bmatrix}, \kappa_{SSHB} = -0.0714\kappa_0, \text{ respectively.}$$

The performance of the carpet cloak with SSHBs comparing with the conventional carpet cloak is demonstrated in Fig. 2(d-i). The near-field and far-field distribution of pressure fields of an object located on the reflecting plane is shown in Fig. 2(d, e), respectively. It is obvious that the presence of the objects disturbs the reflecting wave. In Fig. 2 (f, g), the carpet cloak with SSHBs is used to make the object invisible, whose scattering is identical to conventional carpet cloak that depicted in Fig. 2(hi). All aforementioned simulations, verify the idea of illusionary SHB bump and prove the validity of the carpet cloak with SSHBs.

Afterwards, by applying the compression and complementing transformations (the steps illustrated in Fig. 3), the necessitating homogeneous constitutive parameters are obtained for each region which are given in Table. 1.

| Region 1 | $\bar{\bar{\rho}}_{BA'C'} = \rho_0 \begin{bmatrix} 0.8703 & -0.8978 \\ -0.8978 & 2.0752 \end{bmatrix}, \kappa_{BA'C'} = 2.8441\kappa_0$ |
|---|---|
| | $\bar{\bar{\rho}}_{BK'E} = \rho_0 \begin{bmatrix} 0.6762 & 0.2461 \\ 0.2461 & 1.5685 \end{bmatrix}, \kappa_{BK'E} = 1.4789\kappa_0$ |
| | $\bar{\bar{\rho}}_{BA'K'} = \rho_0 \begin{bmatrix} -10.8272 & 9.1255 \\ 9.1255 & -7.7837 \end{bmatrix}, \kappa_{BA'K'} = -0.1056\kappa_0$ |
| Region 2 | $\bar{\bar{\rho}}_{BMC'} = \rho_0 \begin{bmatrix} 3.6440 & -0.0305 \\ -0.0305 & 0.2747 \end{bmatrix}, \kappa_{BMC'} = 1.4789\kappa_0$ |
| | $\bar{\bar{\rho}}_{ENA'} = \rho_0 \begin{bmatrix} 0.4298 & -0.2741 \\ -0.2741 & 2.5015 \end{bmatrix}, \kappa_{ENA'} = 1.4789\kappa_0$ |
| Region 3 | $\bar{\bar{\rho}}_{GHC'A'} = \rho_0 \begin{bmatrix} -5.3226 & 0.7788 \\ 0.7788 & -0.3018 \end{bmatrix}, \kappa_{GHC'A'} = -1.3897\kappa_0$ |
| Region 4 | $\bar{\bar{\rho}}_{MC'G} = \rho_0 \begin{bmatrix} -1.2121 & -0.3662 \\ -0.3662 & -0.9356 \end{bmatrix}, \kappa_{MC'G} = -1.3897\kappa_0$ |
| | $\bar{\bar{\rho}}_{NA'H} = \rho_0 \begin{bmatrix} -7.6589 & 1.8469 \\ 1.8469 & -0.5759 \end{bmatrix}, \kappa_{NA'H} = -1.3897\kappa_0$ |

**Table 1.** The constitutive materials of the NCAC's regions

Recently, 'acoustic metamaterial technology' have been the subject of numerous researches [44-47]. These studies show promising results for realization of anisotropic media which are presented in Table 1. In Fig. 4, the numerical simulation results are demonstrated to compare the scattering pattern of the object with and without the non-closed carpet cloak. Fig. 4(a, b), display the near-field distribution and far-field scattering pattern of an object with SHB boundaries under the excitation of a Gaussian beam propagating at the angle of $\pi/4$ which disturbed the reflected wave. As demonstrated in Fig. 4(e, f), the presence of designed NCAC neighbor to the object restores the pressure field distribution and far-field pattern the same as the conventional carpet cloak (shown in Fig. 4(c, d)) without any disturbance. The incident fields do not penetrate the hidden region. Because of this feature, the targets shape and its constitutive materials, do not affect on the cloaking behavior of the non-closed structure. The final effect of the device is to hide an arbitrary object without make it blinded. Nevertheless, the simulation results decisively confirm the identical behavior of the designed non-closed carpet cloak with the original carpet cloak. Using homogeneous materials is another supreme benefit that facilitates realization of non-closed invisibilities. It is worth to mention that there is a tradeoff between the extent of the created window on the original cloaking device and difficulty of the realization process. If the cloaking shell is compressed to a very smaller region, it looks more fascinating and is really toward fictions; but the complementary materials take higher values of negative constitutive parameters and the materials' implementation will become harder.

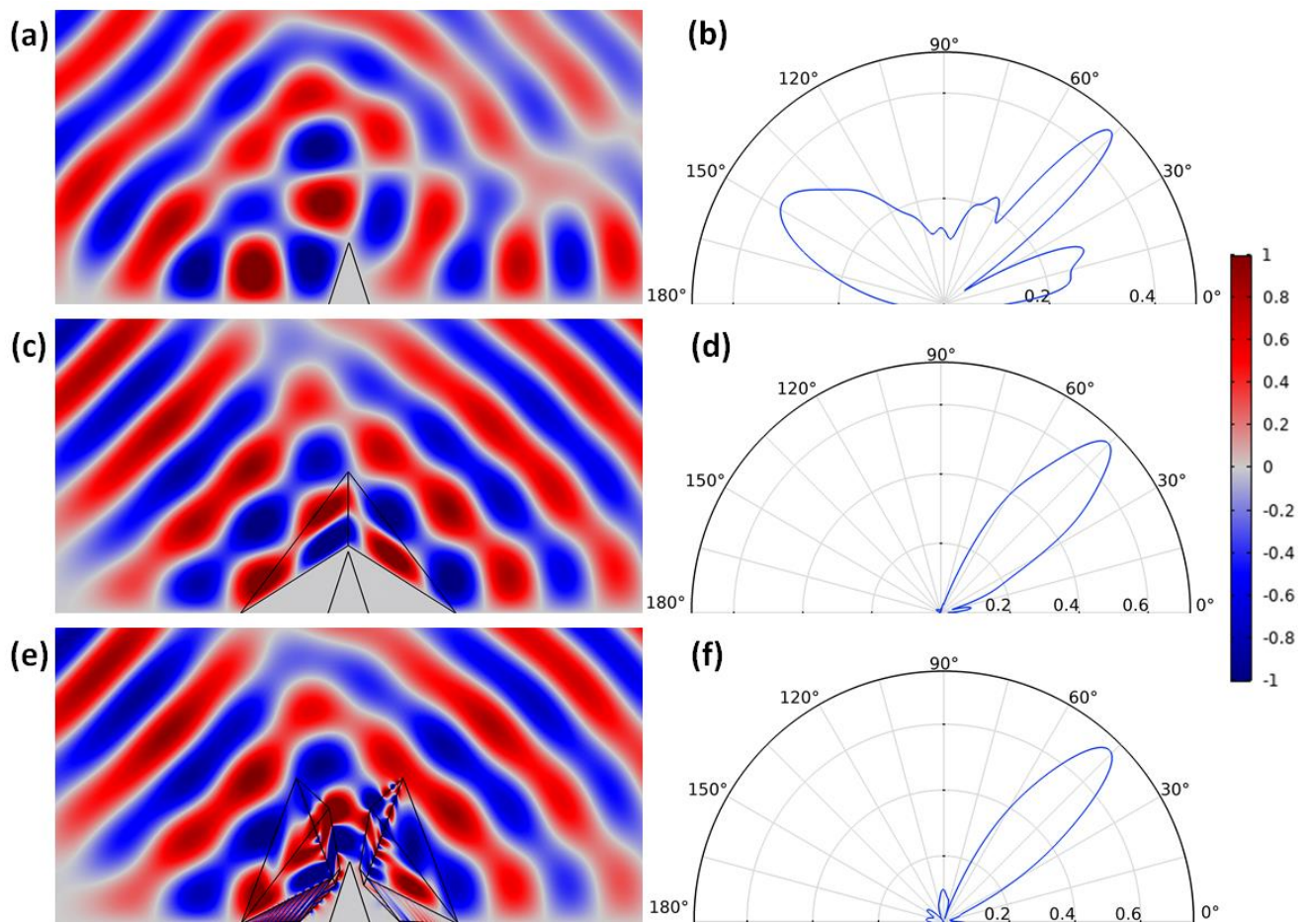

**Figure 4.** Simulation results of non-closed carpet cloak (a-b) Near field distributions and far field scattering pattern of the target with curved geometry and sound hard boundaries that is located on the reflecting plane. The scattered field of reflecting plane is disturbed in the presence of the target (c-d) Near field distributions and far field scattering pattern of the conventional carpet cloak. (e-f) the field distribution of the target near the non-closed carpet cloak. The non-closed device restored the scattering pattern of the object as well as the reflecting plane. It is also similar to field distribution of the conventional structure, which illustrates the validity of the fenestrated carpet cloak.

**Conclusion**

To summarize, we designed and numerically demonstrated a new scheme to acoustic external cloaking by applying sequential step linear coordinate transformations. The cloaking effect of the proposed non-closed device, is independent of objects' architecture because it is based on creating a non-closed hidden region that can make any arbitrary object which is located in, acoustically invisible. Therefore, the target can alter shape or move in hidden region and transform information with outer world without being blinded. There is a tradeoff when the window(s) created on structure are extended which leads to increasing the value of negative constitutive parameters. The presented approach surmounts the inadequacies of the spatially-varying constitutive parameters and dependence of cloaking device to the object. The homogeneous material parameters of the proposed devices, significantly facilitates the realization of acoustic external cloaking devices and opens the way to make the concept of external cloaking applicable in practical level. Due to these benefits, the proposed structures could find applications in varied scenarios such as making submarines invisible from sonar with non-blinded or fenestrate structures. The presented method, also can be applied to other acoustic devices, such as acoustic cavities, waveguides or illusion devices etc. to make them non-closed to outer world that could be useful in future acoustic demands.